\documentclass[aps,pre,twocolumn,groupedaddress,superscriptaddress,showpacs,longbibliography]{revtex4-1}
\usepackage{epsfig,amssymb,amsmath,graphicx,subfigure,hyperref}
\usepackage{tabularx}
\usepackage{float}
\usepackage{setspace}
\usepackage{color}
\usepackage{times}
\usepackage{verbatim}

\usepackage{array} 
\newcommand{\PreserveBackslash}[1]{\let\temp=\\#1\let\\=\temp} \newcolumntype{C}[1]{>{\PreserveBackslash\centering}p{#1}} \newcolumntype{R}[1]{>{\PreserveBackslash\raggedleft}p{#1}} \newcolumntype{L}[1]{>{\PreserveBackslash\raggedright}p{#1}} 
\usepackage{textcomp}

\newcounter{MC}
\newcommand*\MC%
{%
\ifnum\theMC>0%
{MC}\fi%
\ifnum\theMC=0%
{Monte Carlo (MC)\setcounter{MC}{1}}\fi%
}

\newcounter{WHAM}
\newcommand*\WHAM%
{%
\ifnum\theWHAM>0%
{WHAM}\fi%
\ifnum\theWHAM=0%
{the weighted histogram analysis method (WHAM)\setcounter{WHAM}{1}}\fi%
}

\newcounter{fcc}
\newcommand*\fcc%
{%
\ifnum\thefcc>0%
{FCC}\fi%
\ifnum\thefcc=0%
{face-centered cubic (FCC)\setcounter{fcc}{1}}\fi%
}

\newcounter{bcc}
\newcommand*\bcc%
{%
\ifnum\thebcc>0%
{BCC}\fi%
\ifnum\thebcc=0%
{body-centered cubic (BCC)\setcounter{bcc}{1}}\fi%
}

\newcounter{bct}
\newcommand*\bct%
{%
\ifnum\thebct>0%
{BCT}\fi%
\ifnum\thebct=0%
{body-centered tetragonal (BCT)\setcounter{bct}{1}}\fi%
}

\newcounter{hcp}
\newcommand*\hcp%
{%
\ifnum\thehcp>0%
{HCP}\fi%
\ifnum\thehcp=0%
{hexagonally close-packed (HCP)\setcounter{hcp}{1}}\fi%
}

\newcounter{sc}
\newcommand*\scub%
{%
\ifnum\thesc>0%
{SC}\fi%
\ifnum\thesc=0%
{simple cubic (SC)\setcounter{sc}{1}}\fi%
}

\begin{document}

\title{FCC$\leftrightarrow$BCC phase transitions in convex and concave hard particle systems}

\author{Duanduan Wan}
\affiliation{Department of Chemical Engineering, University of Michigan, Ann Arbor, Michigan 48109, USA}
\author{Chrisy Xiyu Du}
\affiliation{Department of Physics, University of Michigan, Ann Arbor, Michigan 48109, USA}
\author{Greg van Anders}
\affiliation{Department of Physics, University of Michigan, Ann Arbor, Michigan 48109, USA}
\affiliation{Department of Physics, Engineering Physics, and Astronomy, Queen's University, Kingston, Ontario, K7L 3N6, Canada}
\author{Sharon C. Glotzer}
\email[E-mail:]{sglotzer@umich.edu}
\affiliation{Department of Chemical Engineering, University of Michigan, Ann Arbor, Michigan 48109, USA}
\affiliation{Department of Physics, University of Michigan, Ann Arbor, Michigan 48109, USA}
\affiliation{Department of Materials Science and Engineering and Biointerfaces Institute, University of Michigan, Ann Arbor, Michigan 48109, USA}

\date{\today}
\begin{abstract}
Particle shape plays an important role in the phase behavior of colloidal self-assembly. Recent progress in particle synthesis has made particles of polyhedral shapes and dimpled spherical shapes available. Here using computer simulations of hard particle models, we study face-centered cubic to body-centered cubic (FCC$\leftrightarrow$BCC) phase transitions in a convex 432 polyhedral shape family and a concave dimpled sphere family. Particles in both families have four-, three-, and two-fold rotational symmetries. Via free energy calculations we find the FCC$\leftrightarrow$BCC transitions in both families are first order. As a previous work reports the FCC$\leftrightarrow$BCC phase transition is first order in a convex 332 family of hard polyhedra, our work provides additional insight into the FCC$\leftrightarrow$BCC transition and how the convexity or concavity of particle shape affects phase transition pathways.  
\end{abstract}

\maketitle

\section{Introduction}

Despite their physical significance, direct observations of solid--solid phase transitions of atomic crystals are difficult as transitions are rapid, and typically occur under extreme conditions and on small length scales. Instead systems at larger length scales, such as colloidal suspensions (e.g., Refs.~\cite{Ise2005,Shevchenko2005,Talapin2009,Travesset2017,Travesset2017_acs}), and micron-sized aqueous droplets (e.g., Refs.~\cite{Bayley2013,Wan2016, Zhang2016}), whose dynamics are significantly slower, provide testbeds for investigating complex, collective phenomena analogous to that in atomic and molecular systems \cite{Gasser2009, Manoharan2015}. Colloidal systems manifesting solid--solid transitions driven by a variety of factors have been explored in experiment (e.g., Refs.~\cite{Yethiraj2004, Zhou2011, Zhang2011, Qi2015, Peng2015, Rossi2015, Mohanty2015}). Among them, one interesting kind of phase transition is that driven by a change in particle shape. Both experimental (e.g., Refs.~\cite{Henzie2012, Young2013, Meijer2017, Gong2017}) and numerical (e.g., Refs.~\cite{Gang2011, Haji-Akbari2009, Kraft2012, Agarwal2011,Damasceno2012_science, Ni2012, Marechal2010, Marechal2010_pre, Gantapara2013, Geng2017, vanAnders2014, vanAnders2014_acs}) studies have shown that particle shape plays an important role in the self-assembled phases of colloidal systems. 

Recent progress in particle synthesis has made many kinds of particle shapes possible. One class of shape is the polyhedron, such as the rhombic dodecahedron \cite{Chen2009, Vutukuri2014, Young2013} and cuboctahedron \cite{Henzie2012, Wang2013}. Intermediate shapes between two polyhedra, with vertex or edge truncation from the bounding shapes, are also available \cite{Chen2009, Chiu2011,  Henzie2012, Wang2013, Young2013, Zheng2014}. Another class of shape is the dimpled sphere, where ``lock-and-key" colloids with a prescibed number of dimples can be synthesized \cite{Sacanna2013, Ivell2013, Odriozola2008, Odriozola2013, Desert2013, Sacanna2010, Wang2013, Ahmed2015}. These two classes of shapes differ in that polyhedral shapes are convex while dimpled spheres are concave (e.g.,~\cite{Sacanna2013, Ivell2013, Odriozola2008, Odriozola2013, Desert2013, Sacanna2010, Wang2013, Ahmed2015}). Despite this difference, shape-driven FCC$\leftrightarrow$BCC phase transitions occur in both systems when particle shape is suitably chosen \cite{Du2017}.   

Using the free energy calculation method developed in Ref.~\cite{Du2017}, here we investigate the FCC$\leftrightarrow$BCC phase transitions in a convex 432 polyhedral shape family \cite{Chen2014, Klotsa2018} and a concave dimpled sphere family \cite{Ahmed2015}, treating particle shape as a thermodynamic variable \cite{vanAnders2015}, to determine the order of each transition. Particles in both families have four-, three-, and two-fold rotational symmetries. Together with the previous report of the 332 polyhedral shape family \cite{Du2017}, in all three cases we find the shape induced  FCC$\leftrightarrow$BCC transition is first order. 

\begin{figure}
\centering 
\includegraphics[width=3.3in]{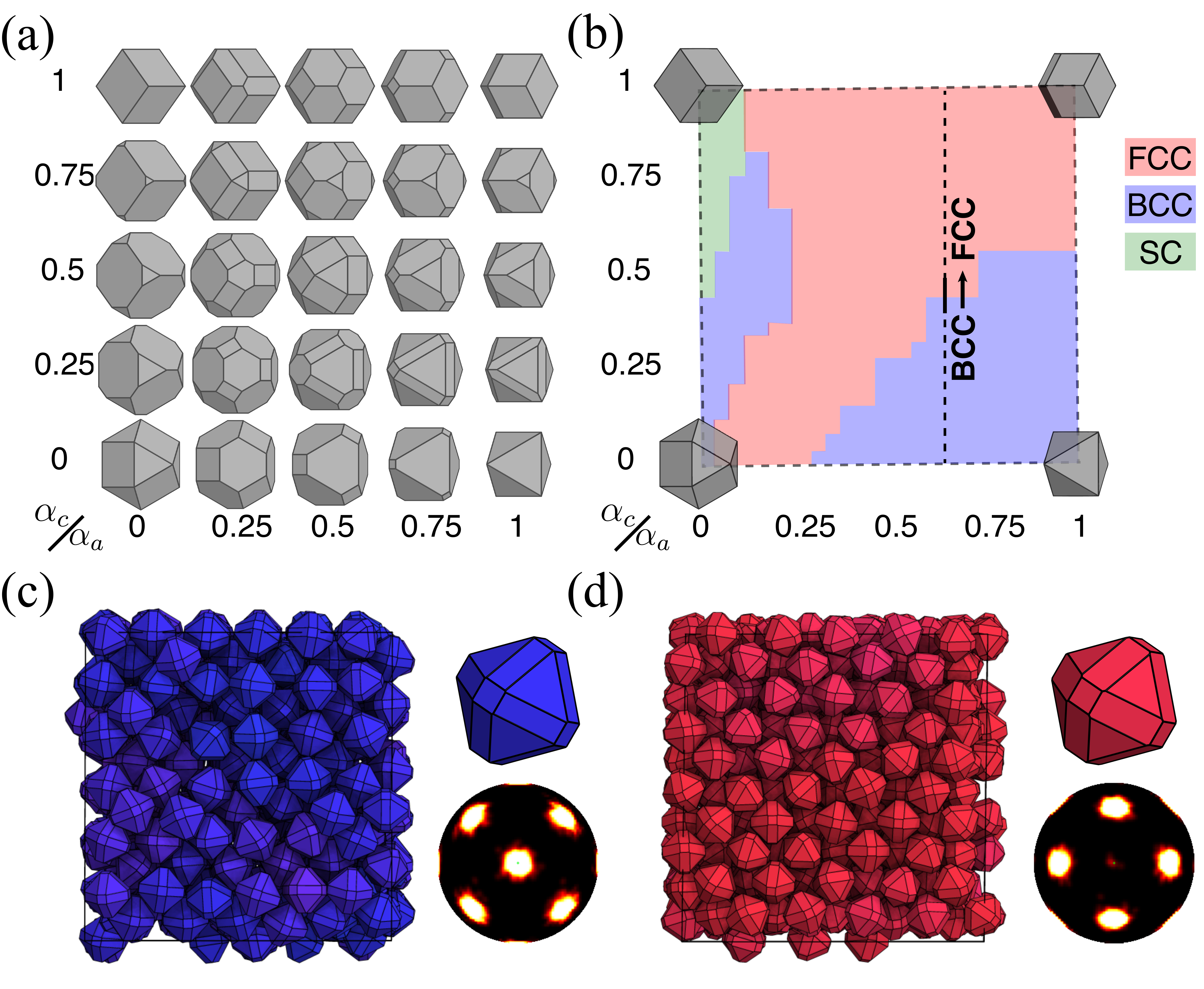} 
\caption{(Color online) (a) Polyhedra in the 432 family as a function of two shape parameters $(\alpha_{a}, \alpha_{c})$. The four shapes at the corners are: Octahedron (1,0), Cuboctahedron (0,0), Rhombic dodecahedron (1,1), Cube (0,1). (b) The rough phase diagram of self-assembled structures of particles in (a) at packing density 0.55 (Replotted from Ref.~\cite{Klotsa2018}). The vertical dashed line at $\alpha_{a}=0.65$ indicates the BCC-to-FCC transition studied in this work. (c-d) Snapshots of the self-assemble structure and the corresponding bond order diagram. (c) $(\alpha_a, \alpha_c) = (0.65, 0.32)$, a BCC structure; (d) $(\alpha_a, \alpha_c) = (0.65, 0.4)$, an FCC structure. }   
\label{432}
\end{figure}

\begin{figure}
\centering 
\includegraphics[width=3.3in]{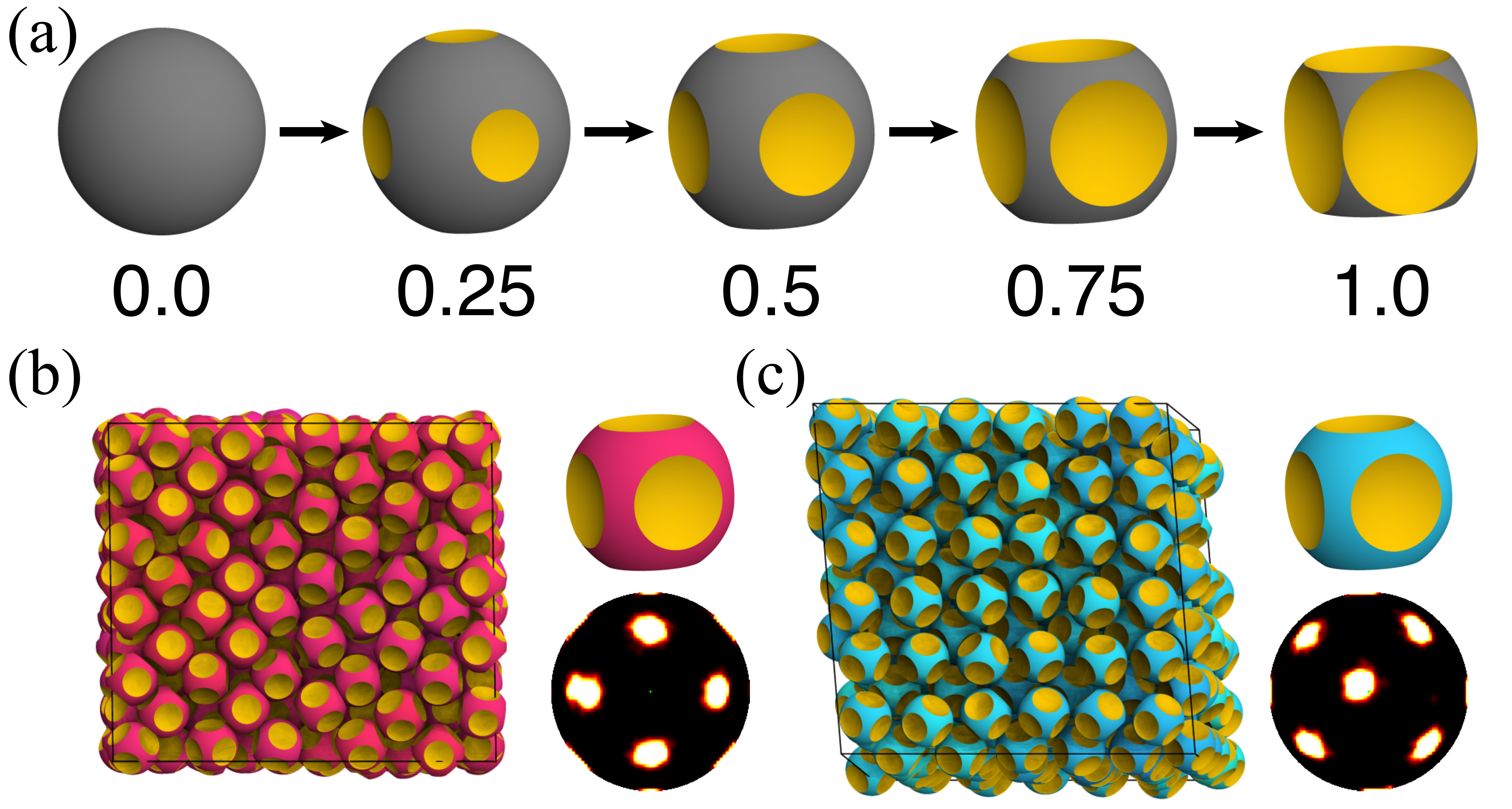} 
\caption{(Color online) (a) Dimpled spheres with $f=0, 0.25, 0.5, 0.75, 1$, respectively. (b-e) Snapshots of the self-assemble structure and the corresponding bond order diagram. (b) $f=0.63$, an FCC structure; (c) $f=0.67$, a sheared BCC structure.}
\label{dimple}
\end{figure}

\section{Models and Methods}

The two families of hard particles we study here can be described using a few shape parameters.  The convex polyhedron shape family with 432-symmetry~\cite{Chen2014} ($\Delta_{432}$) can be described by two shape parameters ($\alpha_a, \alpha_c$), as shown in Fig.~\ref{432}(a).  All the shapes in $\Delta_{432}$ are bound by four shapes: cuboctahedron ($(\alpha_a, \alpha_c) = (0, 0)$), octahedron ($(\alpha_a, \alpha_c) = (1, 0)$), cube ($(\alpha_a, \alpha_c) = (0, 1)$) and rhombic dodecahedron ($(\alpha_a, \alpha_c) = (1, 1)$).  From $(\alpha_a, \alpha_c) = (1, 0)$ to $(0, 0)$, the octahedron has an increasing amount of vertex truncation until $(\alpha_a, \alpha_c) = (0, 0)$, while from $(\alpha_a, \alpha_c) = (1, 0)$ to $(1, 1)$, the octahedron has an increasing amount of edge truncation until $(\alpha_a, \alpha_c) = (1, 1)$.  The same rule applies to other rows and columns in Fig.~\ref{432}(a). In simulations, the volume of particles is rescaled to 1.

For the concave dimpled sphere family~\cite{Ahmed2015}, a dimpled sphere is a spherical cap bounded by the intersection of a central sphere with valence spheres. Here we choose the central and valence spheres of the same radius $r$, with six valence spheres in the $(\pm 1, 0, 0)$,$(0, \pm 1, 0)$, $(0, 0, \pm 1)$ directions.  By choosing these six directions for the valence spheres, we obtain a dimpled sphere with 432-symmetry. The dimpled amount is characterized by the distance $l$ between the central sphere and the valence sphere. When $l=2r$, the central and valence spheres are just touching each other and no dimple is shown; when $l=\sqrt{2}r$, the dimpled sphere has the maximal dimple amount, wherein the two neighboring dimples are touching each other. We thus can use a single shape parameter $f=((2r)^{2}-l^{2})/ (\sqrt{2}r)^{2} \in [0, 1]$ to describe it, where $f=0$ corresponds to no dimple amount and $f=1$ the maximal dimple possible (see examples in Fig.~\ref{dimple}(a)). As above, the volume of particles is rescaled to 1.

\begin{figure}
\centering 
\includegraphics[width=3.3in]{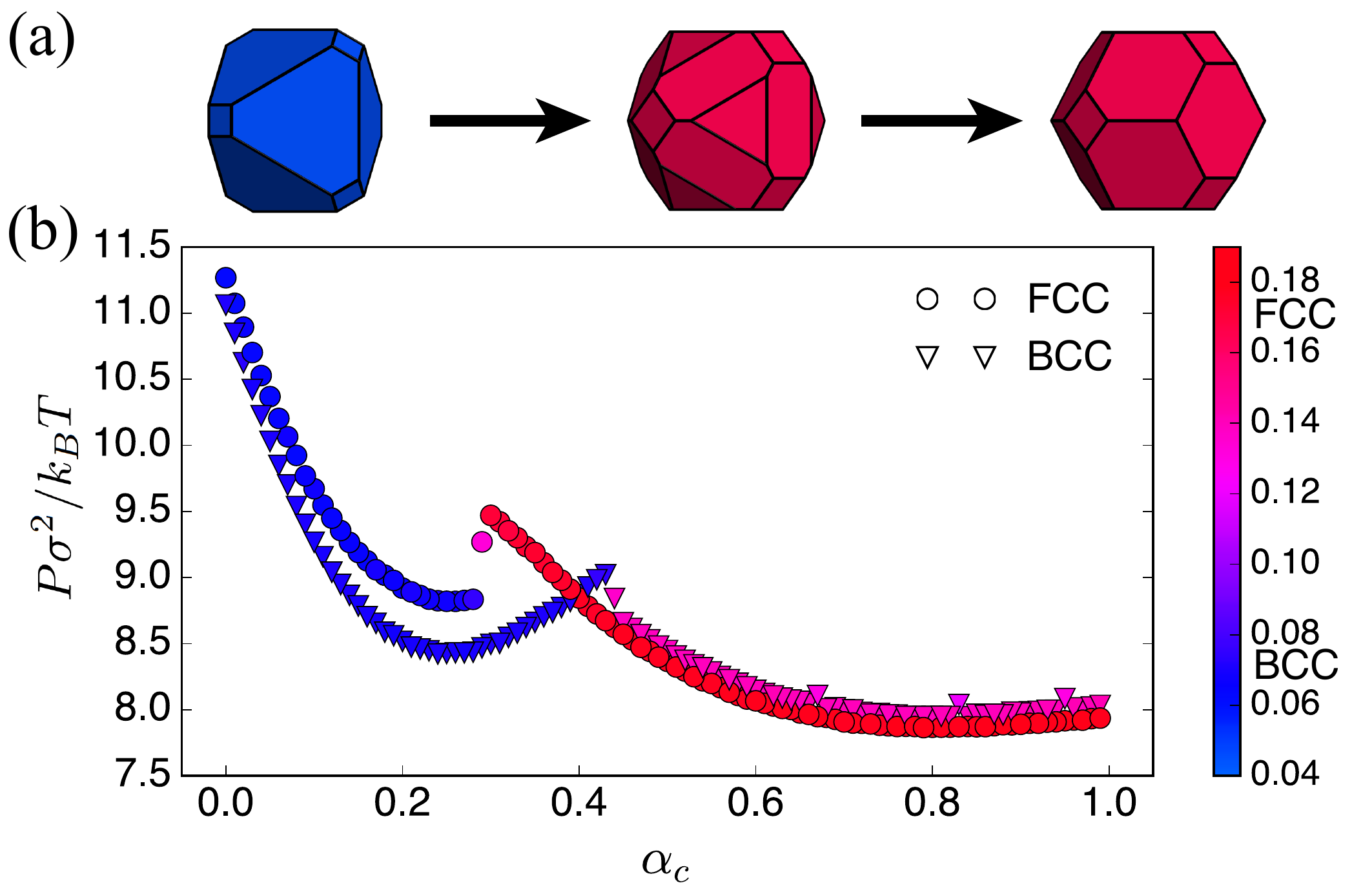} 
\caption{(Color online) (a) Particles with fixed $\alpha_{a}=0.65$ and $\alpha_{c}=0, 0.5, 1$, respectively. (b) Pressures as a function of $\alpha_{c}$ within BCC initialized (triangles) and FCC initialized (circles) systems. Color indicates the value of $\overline{Q_{4}}$ of the final structures of the systems.} 
\label{432_pressure}
\end{figure}

\begin{figure}
\centering 
\includegraphics[width=3.3in]{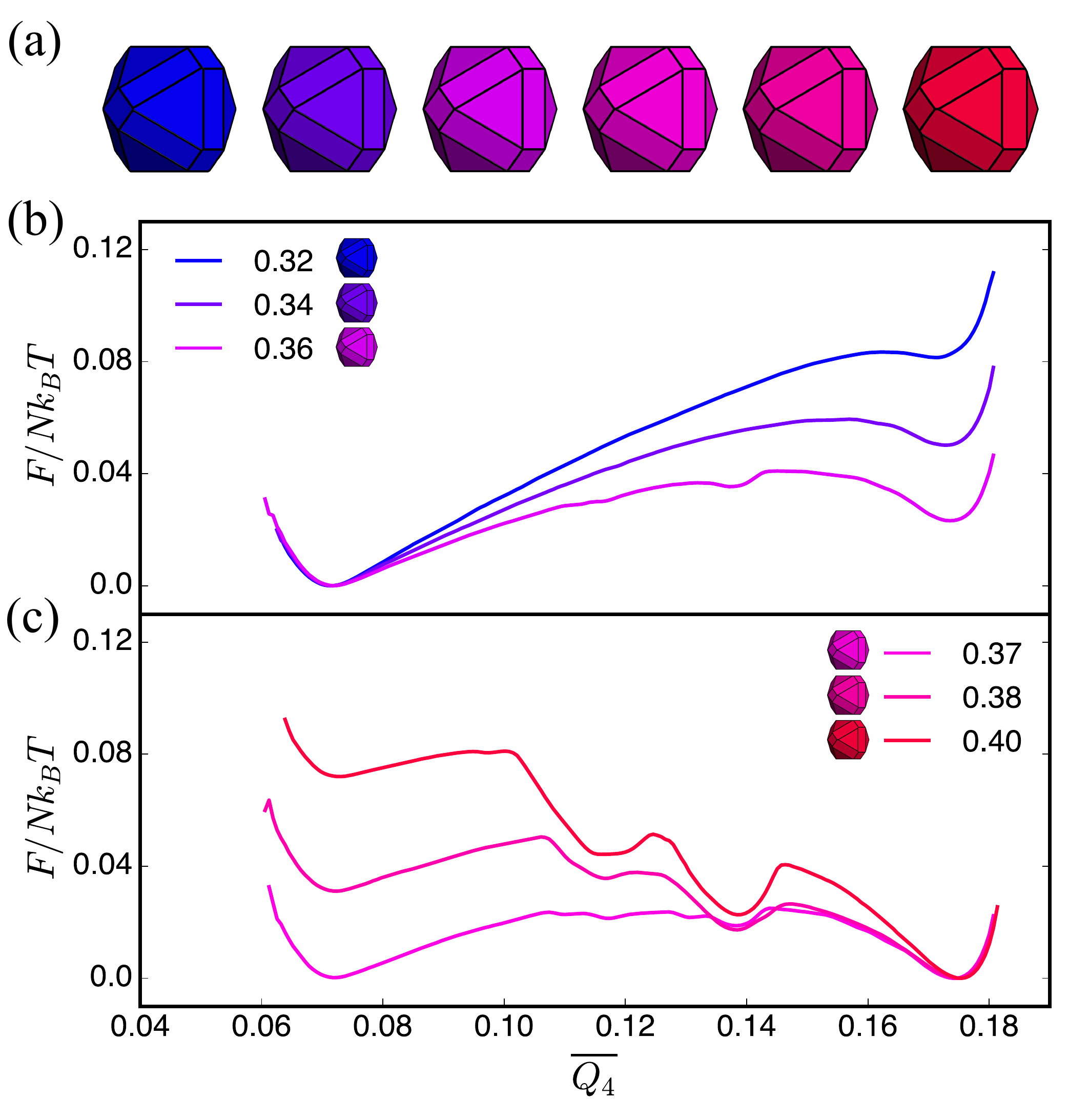} 
\caption{(Color online) (a) Particles used for Landau free energy calculation. (b) Landau free energy curves as a function of order parameter $\overline{Q_{4}}$ for particles with a lower BCC basin. (c) Landau free energy curves as a function of order parameter $\overline{Q_{4}}$ for particles with a lower FCC basin. } 
\label{432_landau}
\end{figure}

To calculate the thermodynamic properties of FCC$\leftrightarrow$BCC transitions in these two systems, we adapt the method in Ref.~\cite{Du2017}. All the simulations are done using the hard particle Monte Carlo (HPMC) module in HOOMD-blue \cite{hoomd,Glaser2015} and data management is done using the signac toolkit \cite{Adorf2018,signacSoftware}. We first calculate the equation of state of the particles in $\Delta_{432}$.  To do so, we initialize systems of $N$ particles in a cubic box with FCC ($N=500$) and BCC ($N=432$) structures and gradually compress the system to reach packing density 0.55. We then relax the system in the $NVT$ ensemble for up to $4\times 10^{7}$ MC sweeps. For each MC sweep, we allow both particle translation and rotation, as well as box shape change (with box volume conserved). The box shape change operation allows the simulation cell to align with the emergent orientation of the crystal~\cite{Filion2009, Graaf2012}. The system is equilibrated for $1.5 \times 10^{7}$ steps and then we begin to measure the pressure. From the pressure measurements we extract an equation of state in terms of pressure versus shape parameter $\alpha_c$, which we use to estimate the location of the solid--solid transitions (see Fig.~\ref{432_pressure}(b)). With this information, we choose sample shapes around the transition area to compute their Landau free energies.  We use the second neighbor-averaged $l=4$ of the spherical harmonic bond order parameter $\overline{Q_{4}}$ \cite{Lechner2008, Steinhart1983} as the order parameter of this system because BCC and FCC structures have two distinct $\overline{Q_{4}}$ values, i.e., BCC has a characteristic value of $\overline{Q_{4}}\approx 0.07$ and FCC has $\overline{Q_{4}}\approx 0.17$ \cite{Du2017}. If the transition happens through continuous intermediate structures, e.g., through a BCT phase, the $\overline{Q_{4}}$ value is expected to change continuously as well. We then compute the Landau free energy for the $\overline{Q_{4}}$ order parameter using umbrella sampling \cite{Kastner2011}.  The spring constant of the biased potential is set to be $k=3.5 \times 10^{4}$ and the window width of $\overline{Q_{4}}$ is 0.004 with $5\times 10^{4}$ samples used for each window.  We use the weighted histogram analysis method \cite{Kumar1992} to reconstruct the free energy curves and the final plot is an average of five independent replica.

\begin{figure}
\centering 
\includegraphics[width=3.3in]{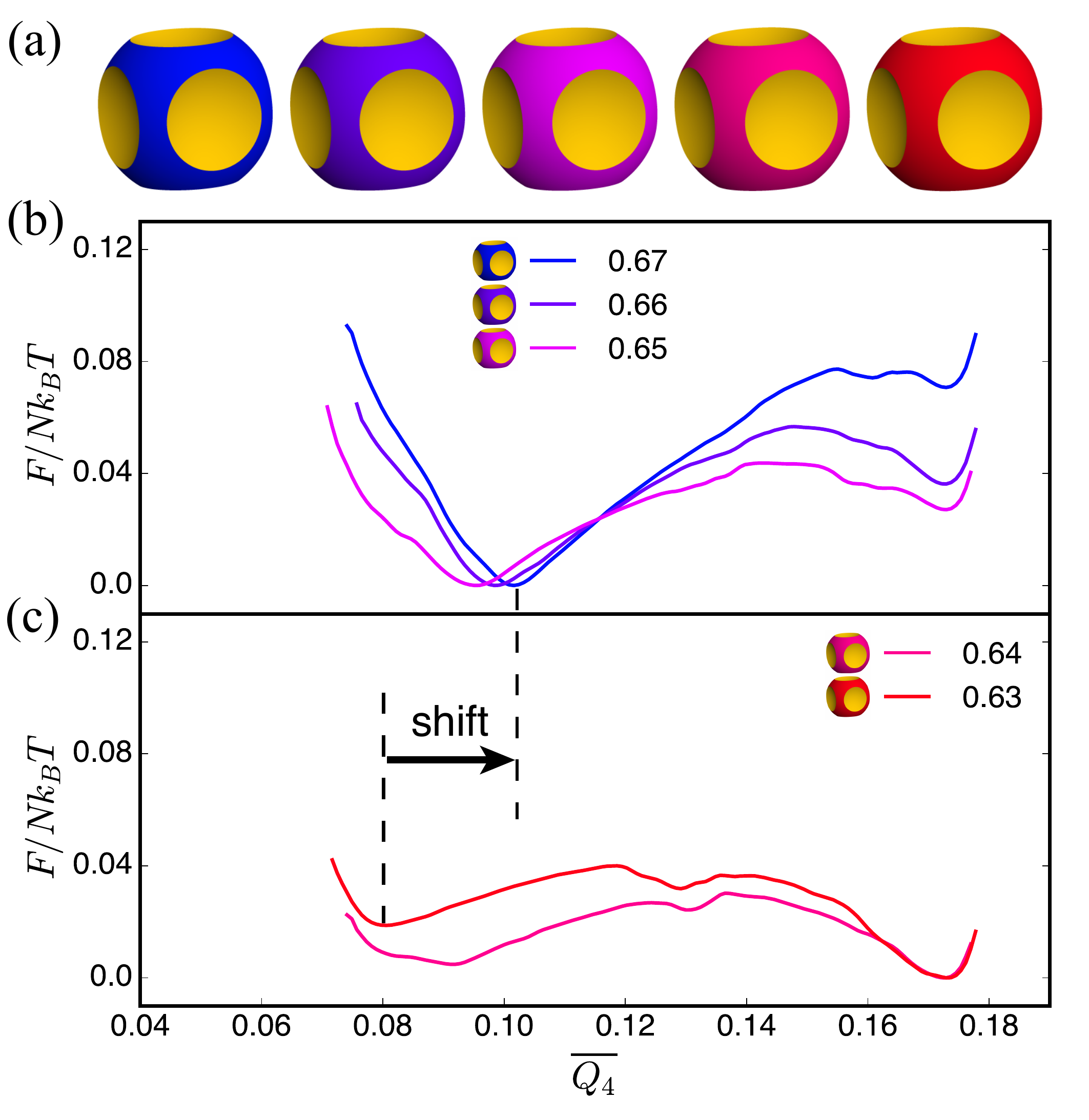} 
\caption{(Color online) (a) Dimpled spheres used for Landau free energy calculation. (b) Landau free energy curves as a function of the order parameter $\overline{Q_{4}}$ for dimpled spheres with a lower BCC basin. (c) Landau free energy curves as a function of the order parameter $\overline{Q_{4}}$ for dimpled spheres with a lower FCC basin. The arrow indicates the shift of the BCC basin. } 
\label{dimple_landau}
\end{figure}

\section{Results and Discussion}

Ref.~\cite{Klotsa2018} studied the self-assembly behavior of particles in $\Delta_{432}$ and Fig.~\ref{432}(b) shows a rough sketch of the three major phases at density 0.55.  Here we are interested in the BCC$\leftrightarrow$FCC transition at fixed $\alpha_{a}=0.65$ and varying $\alpha_{c}$ values (the dashed line in Fig.~\ref{432}(b)). We plot the pressure of the equilibrated systems as a function of $\alpha_{c}$ in Fig.~\ref{432_pressure}(b), with the particles first initialized in BCC and FCC structures, respectively. $\sigma=1$ is the length unit. The color in Fig.~\ref{432_pressure}(b) represents $\overline{Q_{4}}$ of the final structures in the equilibrated systems, with blue for BCC and red for FCC. The color of $\overline{Q_{4}}$ shows that the equilibrated systems are either BCC-like or FCC-like, without any sign of intermediate structures. Pressure curves for both BCC and FCC initialized systems have cusps (in the range about 0.25 to 0.5), which also indicates the transition is first order. Fig.~\ref{432}(c), (d) show snapshots of the equilibrated systems of a BCC and FCC structure, respectively. The structures can be identified from the bond order diagram, which connects a particle with neighboring particles within the first peak of the radial distribution function.  We then calculated the free energy around the value of the shape parameter $\alpha_{c} \approx 0.36$, where the transition takes place. From Fig.~\ref{432_landau}(b), it can be seen that at $\alpha_{c}=0.32$, the system has the lowest free energy in the BCC basin ($\overline{Q_{4}}=0.07$). As $\alpha_{c}$ increases, the minimal free energy of the BCC basin increases while the FCC basin ($\overline{Q_{4}}=0.17$) decreases, which indicates the system begins to prefer an FCC structure. A comparison of the two basins shows that it is a first order transition. The undulations of the $\alpha_{c}=0.38$ and $\alpha_{c}=0.4$ curves in the $0.11-0.15$ range are due to the existence of hexagonally close-packed (HCP) stacking faults in the FCC structure.

We next explore the FCC$\leftrightarrow$BCC transition of the concave dimpled spheres. Because the pressure calculation in HPMC currently does not support concave particles, we identify the phase transition boundary using bond order diagram and $\overline{Q_{4}}$. When $f\lesssim 0.63$, particles tend to self-assemble into an FCC structure; when $f\gtrsim 0.67$, particles tend to self-assemble into a BCC structure. Fig.~\ref{dimple}(b-e) show snapshots of self-assembled structures. The free energy plots in Fig.~\ref{dimple_landau}(b), (c), similar to that in Fig.~\ref{432_landau}(b), (c), have two basins corresponding to the BCC and FCC structures, and show a first order transition as observed in the truncated octahedron system. The BCC basin shifts to the right of $\overline{Q_{4}}\approx 0.07$ and shifts further with increasing $f$ (see the arrow in Fig.~\ref{dimple_landau}(c)) as the BCC structure becomes slightly sheared (Fig.~\ref{dimple}(c)). 

\section{Conclusions and Outlook}
We studied examples of FCC$\leftrightarrow$BCC phase transitions in a convex 432 polyhedral shape family and a concave dimpled sphere family, where shapes in both families have four-, three-, and two-fold rotational symmetries. Together with the previous report on convex 332 polyhedral shape family \cite{Du2017}, in all three cases the FCC$\leftrightarrow$BCC phase transitions are first order. On the other hand, the existence of intermediate BCT structures between FCC and BCC indicates that in Landau theory the BCC$\leftrightarrow$FCC transition could occur via a pair of continuous transitions, e.g., through the Bain pathway \cite{Bain1924}. The Bain pathway has been observed in kinetics of colloidal crystal transformation in experiments (e.g., Ref.~\cite{Casey2012,Weidman2016}). Thus our finding raises the question what factors affect the transition pathway in shape-driven transitions. 
More studies in this direction are encouraged. Furthermore, despite the apparent insensitivity of the overall thermodynamics of the transition to the particle modifications tested here, some discernible differences in the thermodynamics of the transitions were found. Whereas for the convex shapes reported here and in Ref.~\cite{Du2017} there is strong evidence of metastable mixed FCC/HCP stacking developing after the BCC$\leftrightarrow$FCC transition, this was not evident in our study of concave 432-symmetric shapes. This finding indicates that choice of particle shape does afford some control over transition thermodynamics. Understanding the extent to which this is possible will be an important question for future work, given the growing number of examples of shape-shifting colloids that can now be synthesized \cite{Gang2011,Lee2012,Meester2016,Youssef2016, Meijer2017, Gong2017}, the potential for the use of these colloids in developing materials, and the importance of the thermodynamics of solid--solid transitions in determining the viability of these shape-shifting colloids for driving structural reconfiguration~\cite{Solomon2018}.

\acknowledgments  
D.W.\ and C.X.D.\ contributed equally to this work. We thank Brendon Waters for some early work. GvA thanks D.\ Lubensky for helpful conversations. This work was partially supported by a Simons Investigator award from the Simons Foundation to S.C.G. C.X.D. acknowledges support from the University of Michigan Rackham Predoctoral Fellowship Program. Computational resources and services were supported in part by Advanced Research Computing at the University of Michigan, Ann Arbor.

\bibliography{fcc_to_bcc}

\end{document}